\documentclass[conference]{IEEEtran}

\IEEEoverridecommandlockouts
\newtheorem{definition}{Definition}
\usepackage[inline]{enumitem}
\usepackage{multirow}
\usepackage{tikz}
\usepackage{graphicx}
\usetikzlibrary{decorations.markings}
\usepackage{pgfplots}
\usepackage{cite}
\usepackage{amsmath,amssymb,amsfonts}
\usepackage{algorithmic}
\usepackage{graphicx}
\usepackage{textcomp}
\usetikzlibrary{patterns}
\usepackage{xcolor}

\usepackage{booktabs}
\usepgfplotslibrary{external}
\usetikzlibrary{decorations.markings}
\usepackage{caption}
\usepackage{subcaption}
\pgfplotsset{
    legend entry/.initial=,
    every axis plot post/.code={%
        \pgfkeysgetvalue{/pgfplots/legend entry}\tempValue
        \ifx\tempValue\empty
        \else
            \expandafter\addlegendentry\expandafter{\tempValue}%
        \fi
    },
}
\usepgfplotslibrary{groupplots}
\pgfplotsset{compat=1.12}
\usetikzlibrary{patterns}

\def\BibTeX{{\rm B\kern-.05em{\sc i\kern-.025em b}\kern-.08em
    T\kern-.1667em\lower.7ex\hbox{E}\kern-.125emX}}

\usepackage{fancyhdr}
\begin{document}

\title{``When and Where?": Behavior Dominant Location Forecasting with Micro-blog Streams}


\author{\IEEEauthorblockN{Bhaskar Gautam\IEEEauthorrefmark{1},
Annappa Basava\IEEEauthorrefmark{2}, Abhishek Singh\IEEEauthorrefmark{3} and
Amit Agrawal\IEEEauthorrefmark{4}}
\IEEEauthorblockA{Department of Computer Science and Engineering,\\
\IEEEauthorrefmark{1}\IEEEauthorrefmark{2}National Institute of Technology Karnataka, \IEEEauthorrefmark{4}Indian Institute of Technology Roorkee, \IEEEauthorrefmark{3}Amazon\\
Email: $\{$\IEEEauthorrefmark{1} bhaskar.gautam2494,
\IEEEauthorrefmark{3}abhishek5131singh,
\IEEEauthorrefmark{4}amit.agrawal,
\IEEEauthorrefmark{2}annappa$\}$@gmail.com, @ieee.org}}

\maketitle

\begin{abstract}
The proliferation of smartphones and wearable devices has increased the availability of large amounts of geospatial streams to provide significant automated discovery of knowledge in pervasive environments, but most prominent information related to altering interests have not yet adequately capitalized. In this paper, we provide a novel algorithm to exploit the dynamic fluctuations in user's point-of-interest while forecasting the future place of visit with fine granularity. Our proposed algorithm is based on the dynamic formation of collective personality communities using different languages, opinions, geographical and temporal distributions for finding out optimized equivalent content. We performed extensive empirical experiments involving, real-time streams derived from 0.6 million stream tuples of micro-blog comprising 1945 social person fusion with graph algorithm and feed-forward neural network model as a predictive classification model. Lastly, The framework achieves 62.10\% mean average precision on 1,20,000 embeddings on unlabeled users and surprisingly 85.92\% increment on the state-of-the-art approach.
\end{abstract}


\section{Introduction}
\label{sec:Introduction}
The branch of knowledge discovery and information science is expanding with the expansions of data stream velocity. This acceleration arise the need to fully-automated recommendation model to facilitate a reproducible framework having wide applicability to online social network streams. Initially, these on stream networks facilitate individuals to connect and communicate with peer network users but recent advancements towards location-based social network simplifies the low-cost self-position reporting on contemporary networks. For example, a recent study proved that exploiting progression of tuples provide 88\% more exposure compared to the personal recommendations and will have the capability to leverage 77\% of smartphone users check-ins stream into personalized promotional business~\cite{30}. However increase in usage of networks among geographical distributed social media users the streams comprising self-position reporting are usually submerged with the desire and intention of users. Hence, reliable automated retrieval methods of classification are needed to correlate the fluctuating interest of discussed users with the reporting locations.

In this paper, we focus on fully automated annotation and in-depth recommendation of the plausible place of visit to the social user. Interestingly, we reduce dependency threshold on the number of tweets per day comprising self-location reporting count while assuming the most recent tweet reflecting the personality of the user. These long range and in-depth recommendations are helpful in personalized advertisements, campaigns, and peer-to-peer trajectory promotions. There are lot of conventional approaches~\cite{36,37} towards removal of cold start problem while imposing strict restrictions on check-ins distribution and avoiding users with sparse history. This feature-rich dimensions available on the social media and location-based social network motivate us to exploit fluctuating behavior of user towards long-range recommendation. However, location reporting tuples are merged with redundancy and short text languages used in micro-blogs. We address these gaps in the paper. We propose novel \textit{Forecasting Spot} framework for recognition of self-location reporting and in-depth long range recommending plausible spots. Specifically, we summarize our contribution to the following:
\begin{enumerate}
\item We provide a formal definition, search, normalization and validation methods for recognizing communities on microblog streams. Later form communities of social users while exploiting their fluctuating interest that resembles with self-location reporting stream tuples.
\item We provide a reliable automated approach to annotate stream tuples with self-location reporting categories.
\item We propose a novel \textit{Forecasting Spot} framework running on top of a multilayer feed-forward neural network to recommend a plausible spot for the individual user. These long-range predictions support over weekdays with in-depth granularity over time.
\item We provide a brief summary of the implications that we observed during experiment analysis of framework potentially for personality modeling of a social user. Later we provide the pseudo methodology to replicate the results of the entire paper.
\end{enumerate}
\noindent The remainder of this paper is structured as follows: In Section \ref{sec:related_work}, we highlight the prominent work carried out so far along with the encountered research gaps, Section~\ref{sec:notation_definition} describes the definitions of individual communities and notation used in resp of the paper. Section~\ref{sec:comm_search_method} and~\ref{sec:comm_exploitation} delineate the community search followed by community exploitation methods respectively. Section~\ref{sec:prop_model} describes the methodology followed in the formation of the entire prediction model. Section~\ref{sec:exp_eval} depicts the evaluation and result while comparing with state-of-the-art contributions. Finally, the paper concludes in Section~\ref{sec:conclusion} with a discussion for future contributions from authors.

\section{Related Work}
\label{sec:related_work}
Youcef et al.~\cite{37} present tribase algorithm for community extraction on the social network corpus and demonstrated structural characteristics of the cluster. The Qianyi et al.~\cite{38} propose spectral clustering and low-rank matrix factorization method to determine community detection for emerging social networks. They compare the efficacy over different adjustment while using the Davies-Bouldin index and silhouette index as an evaluation metrics. The model proposed by Nagarajan et al.~\cite{39} are based on link-content and Gibbs sampling algorithm to recommend friends in the social network. The Huy et al.~\cite{40} demonstrated the first influencer model to predict the influence of diffusion in communities. Most other studies showed that spatiotemporal characteristics have an impact on communities information diffusion and properties of different communities. Based on the exploitation of historical tweets and personality features, the Arun et al.~\cite{41} proposed methodology for predicting the next visit of the user. Their work is more inclined towards exploitation of \textit{Latent Dirichlet Allocation} and \textit{Binary Relevance} methods while considering radius and time as an additional feature. The Jennifer et al.~\cite{42} exploit the microblog content with the linguistic inquiry and word count tool and MRC psycholinguistic database to obtain linguistic and psycholinguistic dimensions. Lastly, the \textit{Gaussian process} and \textit{ZeroR} algorithms are used to estimate score for each feature. However, there is a specific need of a framework which is widely applicable to streams different network whom should be independent of geotagged content and capable to exploit fluctuated point of interest on these networks.

\section{Notations and Definition}
\label{sec:notation_definition}
\subsection{Community}
\begin{definition}
A community in microblog stream comprised of a group of the likely minded person sharing a common point of interest. A fusion of these communities with a location-based social network turns into a feature-rich network. It is majorly comprised of four widely available dimensions labeled as - temporal, opinion, language, and geographical attributes. The pattern of these spans are able to recognize periodicity by focusing on a detail level of demographics.
\end{definition}

\subsection{Community Resolution}
\begin{definition}
Given an identity $M_{\alpha}^{A}$ of user $\alpha$ on location based social network $\lambda_A$, find her correct community, C on contemporary network provides that shares common semantics and logic structures.
\begin{displaymath}
M_{\alpha}^{A} \rightarrow \{ Community, C\}
\end{displaymath}
\end{definition}
\emph{Generic Methodology:} The process of community resolution in location-based social networks follow two subprocess - \textit{community search} and \textit{community validation}. Community search process list a set of social users on $\lambda_A$, which pose similar point-of-interest to a given $\alpha$ users on $\lambda_A$ and possibly belong to community, C. Community validation process then evaluates the similarity score between $M_{\alpha}^{A}$ with every candidate identity returned by community search process on defined metrics. Candidate identities are then ranked on the basis of similarity score and the candidate identities with the highest score are eligible to form a community, C with $M_{\alpha}^{A}$.

\subsection{Community Search}
\begin{definition}
For a user $\alpha$, given her identity $M_{\alpha}^{A}$ on location based social network $\lambda_A$ and a search parameter Q, find a set of identities $M_{\beta}^{j}$ on contemporary network $\lambda_A$ such that Q($M_{\alpha}^{A}$) $\simeq$ $M_{\beta}^{j}$ which resembles similar point of interests. Each member in the candidate set $\{M_{\alpha}^{A},\ Q\}$ termed as candidate identity whose cardinality is measured with N.
\begin{displaymath}
\{M_{\alpha}^{A},\ Q\} \rightarrow \{M_{\beta}^{1},..,M_{\beta}^{j},..,M_{\beta}^{N}\} 
\end{displaymath}
\end{definition}

\emph{Generic Method}: Any search method takes a source and a set of search parameter as input to retrieve a set of potential candidate items. For an identity search algorithm, the source can be identities on social network $\lambda_A$ and a search parameter based on dynamic variations to temporal, opinion, language, and geographical attributes. In return for closely associated candidates with the search parameter.

\subsection{Community Validation}
\begin{definition}
Given a set of candidate identities defined to be \{$M_{\alpha1}^{A},\cdots,M_{\alpha j}^{A},\cdots,M_{\alpha N}^{A}$\} on location based social network $\lambda_A$, a search parameter Q and a similarity function S, locate an identity pair ($M_{\alpha j}^{A}$, $M_{\alpha N}^{A}$) such that S($M_{\alpha}^{A}$) = $\max\big\{S(Q, M_{\alpha1}^{A}),\cdots,S(Q, M_{\alpha j}^{A}),\cdots,S(Q, M_{\alpha N}^{A})\big\}$, max is the highest similarity score and cardinality threshold that are eligible to form a community with parameter Q.
\begin{displaymath}
\{M_{\alpha}^{A},\ Q,\ S\} \rightarrow \{M_{\alpha j}^{A},\ M_{\alpha N}^{A}\} \rightarrow Community,\ C
\end{displaymath}
\end{definition}
\emph{Generic Method}: A community validation algorithm identifies the correspondence between search parameter Q and each candidate identity by $M_{\alpha N}^{A}$ by evaluating a similarity threshold S($M_{\alpha}^{A}$) and rank the candidate identity set on the basis of similarity score. Candidate identity pair ($M_{\alpha j}^{A}$, $M_{\alpha N}^{A}$) with highest match score is eligible to form Community, C. Similarity score can be calculated by methods as opinion similarity method, temporal similarity method, geographical network similarity method, language similarity method while leveraging meta-attributes such as location, timestamp, post and language. These methods formulates the similarity using \textit{levenshtein distance} algorithm.

\subsection{Community Normalizing}
\begin{definition}
Given a set of candidate identities P defined to be \{$M_{\alpha1}^{A},\cdots,M_{\alpha j}^{A},\cdots,M_{\alpha N}^{A}$\} which are eligible to form a community on location based social network $\lambda_A$, find an optimal identities set \{$M_{\alpha j}^{A}$, $M_{\alpha N}^{A}$\} $\subseteq$ P representing a set of quality attribute candidates.
 \begin{displaymath}
 C_\alpha^{A} = \big\{ \sum_{i=0}^{\infty} \lbrace\,M_{\alpha N}^{i}\,\rbrace \geqslant \emptyset\, | \, \, \forall \, (M_{\alpha}^{A}) \in \lambda_A \big\}
 \end{displaymath} 
\end{definition}
\emph{Generic Method}: The normalizing community method ensures the presence of an equal contribution of reporting domains competing in the community. The algorithm maintains a threshold based sanity assertion on the cardinality of eligible candidate identities set and domains of reporting locations. The method balances these domains on prioritizing the user based on a temporal attribute of the contemporary micro-blog post.

\section{Community Search Methods}
\label{sec:comm_search_method}
In this section, we discuss the community search methods proposed to search for a set of identities having a similar point of interests on Twitter text streams. The community resolution methods are - \textit{Dynamic Content Opinion Search}, \textit{Dynamic Language Opinion Search}, \textit{Dynamic Location Network Search} and \textit{Dynamic Timestamp Opinion Search}. These methods exploited text streams meta-data of the user publicly available on microblog social network and applied to the user on location-based social network text streams as well. We exploit the least possible information while ensuring the existence of attributes in district text streams of the user in comparison to the state-of-the-art community algorithms~\cite{3} which exploit detailed meta-data of the user. We now elaborate each of the community search methods in detail.

\begin{table*}[!ht]
\centering
\caption{Classification of dynamic language communities based on continents and native language codes}
\label{fig:Lang_Community_Class}
\begin{tabular}{llcc}
\hline
\textbf{Communities} & \textbf{Native Languages Codes} & \textbf{CSL Index} & \textbf{Distribution (\%)}\\ \hline \hline
Africa              & PT, ES, ZA, EU                              & 0.0024 & 10.25\\
Antarctica          & EN                              & 0.0005  & 2.56\\
Asia                & ID, TL, TR, VI, ZH, HI, RU, TH, AR, KO, HE, JA                              & 0.0069 & 30.76\\
Australia           & EN                              & 0.0005  & 2.56\\
Europe              & DA, ES, FR, PL, ET, DE, TR, PT, SV, CS, NL, SL, EL, &  0.0146 & 66.66\\
					& NO, IT, LV, RO, CY, HU, IS, RU, UK, EU, LT, BG, FI                             &   & \\
North America       & HT                              & 0.0011  & 2.56\\
South America       & EN                              & 0.0005  & 2.56 \\ \hline
\end{tabular}
\end{table*}

\subsection{Dynamic Opinions Search}
\begin{figure}[!ht]
\centering
\includegraphics[scale=0.66]{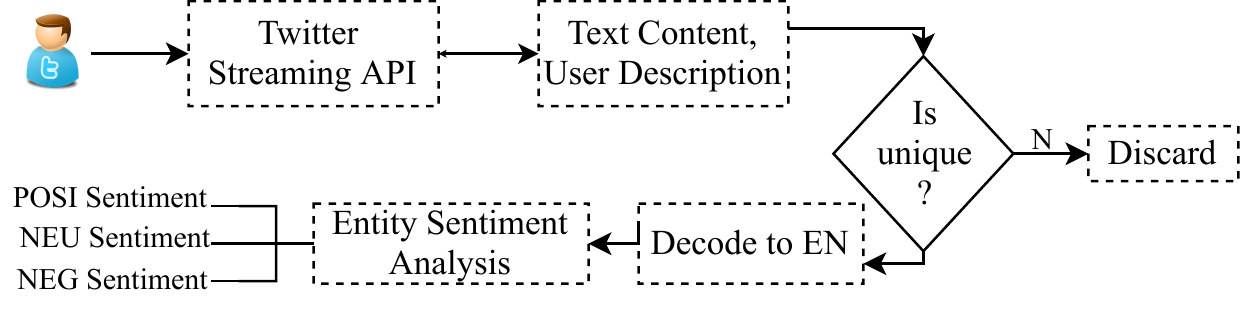}
\caption{Dynamic Opinions pseudo algorithm.}
\label{fig:Content_Opinion}
\end{figure}

A text streams of social media network is generally comprised of textual content that user-author, modified author, and corresponding brief description of user identity. Owing to the popularity of location influenced network and ways to gather affected social networks together, a user is expedited with a choice to push check-ins leveraged content on contemporary networks simultaneously. For example, Foursquare provides functionality to interconnect its network with Twitter snowflake to push user streams on Twitter timeline interlinked with foursquare simultaneously. Since the opinion of a user keeps changing with the change in time and place~\cite{5} respectively. Hence, it is more viable that whenever a user visits a location, it generates the content influenced by its sentiments at time instance and may differ after visiting the contemporary place. Such sentiment behavior of the user can be exposed by Twitter streaming API which provides the ``timestamp'' along tweet, i.e., based on the timezone a timestamp is generated while pushing every tweet. The timestamp can be exploited to segregate both users whose (1) sentiments remain consistent on visiting a location, and (2) whose sentiment differs once visiting a place. In this paper, we focus on both categories of text streams irrespective of change in sentiments to embrace a diversity of different users. However, we plan to use this classification for our future contribution. The search ends up with the formation of three communities that are having a similar engagement for a fixed time quanta. The flow of opinion search algorithm is described in Figure~\ref{fig:Content_Opinion}.
To each tuple as described in Section~\ref{sec:corpus_overview}, we enrich the author profile content and textual attributes (tweet) by normalizing them to lowercase and removing any non-ASCII characters. We impose minimum processing limit to three length characters to sustain quality streams. For determining the uniqueness among textual attributes, we prefer state-of-the-art \textit{Levenshtein Distance} among more apparent candidates in string similarity metrics due to the presence of short text languages. We ensure uniformity in determining dynamic influenced features while translating contents into the English language. Lastly, our proposed algorithm ends up with the formation of three communities using Entity Sentiment Analysis~\cite{7} based on magnitude ($\tau_1$) and sentiment score ($\tau_2$). Although in-depth classification can also be possible, instead we restrict to communities termed as - \textit{Positive Sentiments} having $\tau_1$ and $\tau_2$ greater than 0.0 respectively, \textit{Neutral Sentiments} having $\tau_1\leftarrow0.0$ and $\tau_2\geq0.0$ lastly, \textit{Negative Sentiments} with $\tau_1$ and $\tau_2$ greater and less than 0.0 respectively.

\subsection{Dynamic Language Search}
This method exploited a like-minded users having similar native language and assumed that users having similarity in first spoken languages tends to visit the same place. We formed communities based on widely used dialects~\cite{12} on microblog streams. Although, the language diversity on the Twitter social network is extensively studied by authors~\cite{10,13} while forming local communities on a country level but as an extension we contribute towards the formation of global communities at continent level. For example, there are around 7,097 macro-languages spoken approximately by 6.7 billion peoples belonging to different countries as first language preference~\cite{15}. However, we exploit only 39 languages and countries where these languages expressed as the primary language
\begin{equation}
\resizebox{.85\hsize}{!}{ $CSL\ Index = \frac{Count\ of\ Languages\ Used\ in\ Tuple}{Distinct\ Users\ available\ in\ Community}$} \label{eq:stci}
\end{equation}
These mapped countries are later classified into corresponding continents ensuring that if dialect is spoken natively in more than two nations, then it belongs to contemporary communities as well. The contemporary languages termed as - \textit{Danish (DA)}, \textit{French (FR)}, \textit{Polish (PL)}, \textit{Estonian (ET)}, \textit{German (DE)}, \textit{Swedish (SV)}, \textit{Czech (CS)}, \textit{Dutch (NL)}, \textit{Norwegian (NO)}, \textit{Italian (IT)}, \textit{Latvian (LV)}, \textit{Romanian (RO)}, \textit{Hungarian (HU)}, \textit{Russian (RU)}, \textit{Ukrainian (UK)}, \textit{Basque (EU)}, \textit{Lithuanian (LT)}, \textit{Bulgarian (BG)}, \textit{Finnish (FI)}, \textit{Slovenian (SL)}, \textit{Greek (EL)}, \textit{Indonesian (ID)}, \textit{Tagalog (TL)}, \textit{Turkish (TR)}, \textit{Vietnamese (VI)}, \textit{Chinese (ZH)}, \textit{Hindi (HI)}, \textit{Russian (RU)}, \textit{Thai (TH)}, \textit{Arabic (AR)}, \textit{Korean (KO)}, \textit{Hebrew (HE)}, \textit{Japanese (JA)}, \textit{Haitian	(HT)}, \textit{Portuguese (PT)}, \textit{Spanish (ES)}, \textit{Zulu (ZU)}, \textit{Basque (EU)} and \textit{English (EN)} which were mapped into seven continents. It might be interesting to evaluate the contribution of languages in a community stream and distribution of representative dialects. We term this assessment as Community Stream Language (CSL) Index and keep track on a fraction of languages used by tuples in a stream relative to count of distinct users encountered as shown in Equation~\ref{eq:stci} and become more diverse with an increment of the index. The summary of an entire classification delineated into Table~\ref{fig:Lang_Community_Class} comprising a distribution of each language in representative community and CSL Index.

\subsection{Dynamic Geographical Network Search}
A network of self-reported positioning in micro-blogs is widely explored paradigm of social users and was introduced thoroughly by Watts et al.~\cite{23}. The method exploits the self-check-in attribute of the social user with an assumption that conformity in self-position reporting to social network users are always influenced towards individual personality. The self-check-in meta attribute of a tuple is always defined with the involvement of lawful owner perception towards the location entity apart from itself as compared to other meta-attributes where dependency does not exist ever. If any social user reports their visibility to its timeline, then it is more evident for recommendations. The self-reporting streams from micro-blogs comprising tuples inconsistent timestamp difference while comparing to continuous streams of trajectories from the monitoring device. The self-reporting network algorithm explores the possibility of community in erratic tuple streams.
\begin{figure}[ht!]
\centering
\includegraphics[scale=0.21]{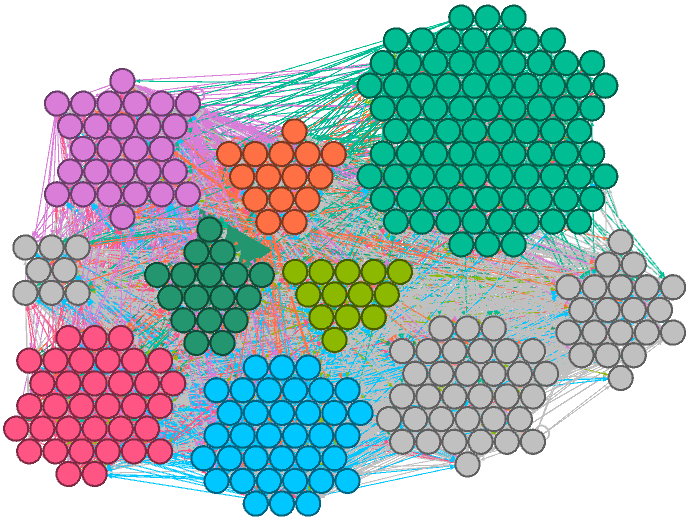}
\caption{Circular packing representation of clustering algorithm forming ten communities of self-reporting positions}
\label{fig:Location_Opinion}
\end{figure}
The state-of-the-art clustering algorithm demands the proper tuning of parameter due to change in performance when corpus varies. For example, the recommended parameter value of \textit{maxD}, \textit{K}, \textit{minPts} and \textit{deltaT} be 140 meters, top 25 correlations, 7 minimum number of points and 7 days respectively for Pande et al. algorithm~\cite{18} on Chicago corpus to achieve efficient clustering. Hence towards automation, we exploit Vincent et al. algorithm~\cite{21} for clustering of dynamically influenced communities comprising self-reporting positions with microblog streams. The algorithm ends up with the formation of 10 communities as shown in Figure~\ref{fig:Location_Opinion} for corpus described in Section~\ref{sec:corpus_overview}. The majorly influenced nations are indexed as - \textit{South Africa}, \textit{Mozambique}, \textit{New Zealand}, \textit{Australia}, \textit{India}, \textit{China}, \textit{Brazil}, \textit{Uruguay}, \textit{Argentina}, \textit{USA}, \textit{Canada}, \textit{Mexico}, \textit{British Columbia}, \textit{Northwest Territories}, \textit{Russia} and \textit{Kazakhstan}.

\subsection{Dynamic Temporal Opinions Search}
This method exploits a like-minded user tends to visit the location at specific and specified fixed instance of time quanta and was influenced by Hossein et al.~\cite{9}. For example, two users $\alpha_1$ and $\alpha_2$ push their intention about location into their timeline but correspondingly at the same time. However, it might be interesting when they share almost similar perception with the contemporary place. This method explores timestamp with contextual attributes of a tuple and assumes the absence of related textual content amongst tuples having a different timestamp.
\begin{figure}[h!]
\centering
\includegraphics[scale=0.66]{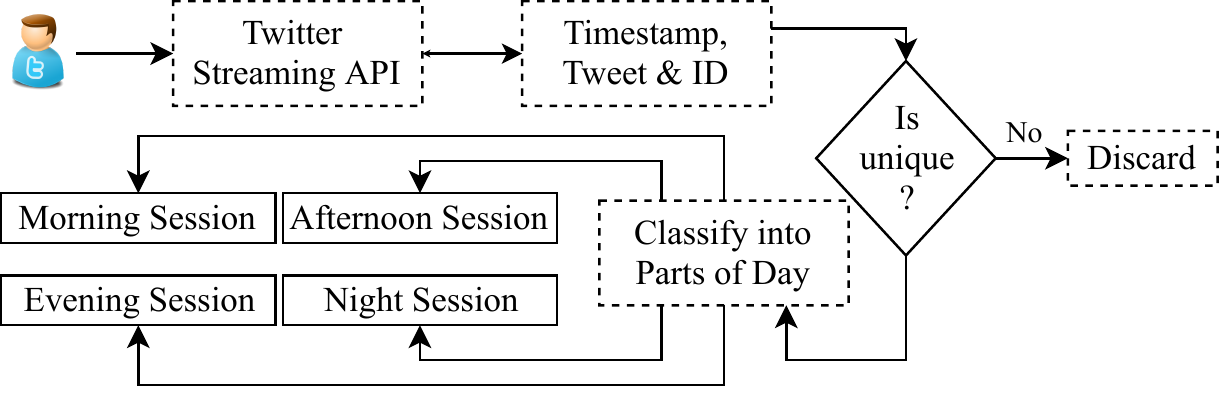}
\caption{Dynamic Temporal Opinions pseudo algorithm.}
\label{fig:Timestamp_Opinion}
\end{figure}
The flow of opinion with time search algorithm is described in Figure~\ref{fig:Timestamp_Opinion}. To each tuple of streams described in Section~\ref{sec:corpus_overview}, we first normalize into lowercase after removing any non-ASCII characters. As discussed earlier, we prefer state-of-the-art \textit{Levenshtein Distance} among more apparent candidates in string similarity metrics to ensure the uniformity among tuples with similar content but having a different timestamp. Lastly, our proposed algorithm ends up with the formation of four communities based on the state-of-the-art description of parts of the day. We termed these communities as - \textit{Morning Session} specified from 12 to 11:59 am, \textit{Afternoon Session} specified from 12 to 16 pm, \textit{Evening Session} specified from 16 to 20 pm and \textit{Night Session} specified from 20 to 23:59 pm.

\section{Community Exploitation}
\label{sec:comm_exploitation}
Given normalized community streams of like-minded users, we use the following methods to recommend the future place of visit for the social user - Personality Modeling and Location Advisor. We then assign dynamic priorities individually to advised locations. The aim of prioritizing these planned locations is to retrieve correct forecasting of locations having high confidence score.

\textbf{Personality Modeling}: We exploit psychological features of the individual community to correlate their self-reporting locations with personality traits of the user. The \textit{Openness to experience}, \textit{Conscientiousness}, \textit{Emotional Range}, \textit{Agreeableness}, \textit{Extraversion}, \textit{Needs}, \textit{motivating factors values}  and aspects of \textit{human needs} characteristics of individuals are explored using state-of-the-art personality profiling model~\cite{25} based on global vectors for word representation. Since they were differently associated with user's life and characteristics of geographical distribution. These characteristic values are comprised of numerical values which are later normalized into a scale ranging from 0 to 100.

\textbf{Location Advisor}: The methods initiate with the formation of distributed embedding representation for each variable length tuple of a social network with Quoc et al.~\cite{33} approach. Simultaneously, we transform the individual timestamp of an incoming tweet into two hot encoding representation vector for weekday and hour respectively. Later, we correlate these vectors space embeddings with the \textit{personality modeling} embeddings as discussed in Section~\ref{sec:comm_exploitation}. Finally, we have unique 109 length vector space embedding and is considers as representative of self-location reporting class assigned to tweet. We manually remove the similar tweet from a corpus using \textit{Levenshtein Distance} while leaving the task of automatically identification of unique tuples among streams on our future work using Marc et al.~\cite{34} approach. The embeddings will become input to feed-forward artificial neural network model having 900 neutrons per hidden layer running up to 300 iterations with ``Adam'' and ``ReLu'' as a solver and activation function. We also proposed distributed representation of each word embedding comprising of variable length social network tuple to evaluate the efficacy of our framework using Tomas et al.~\cite{35} approach. Later, the method returns the estimated spot from the community having maximum confidence score (prediction probability).

We compare the efficiency of these vector space algorithmic models for a geo-annotated corpus with variation in the type of input feature provided to them. This contemporary input features along with algorithmic model are indexed below: \begin{enumerate*}
\item personality traits, weekdays, hours and distributed embedding for each variable, \item personality traits, weekdays, hours and distributed representation of each word embedding, \item weekdays, hours and distributed representation of each word embedding and \item weekdays, hours and distributed representation of each variable \end{enumerate*}. We evaluate the output layer efficacy of each proposed model using standard multi-label classification algorithms. The tunning parameters of each algorithmic models are as follows \begin{enumerate*} \item Support Vector classification uses Gaussian radial basis function kernel with 0.10 epsilon.\item Logistics Regression uses linear least squares with l2 regularization method\item Random forest classifier has an upper bound of 50 on the maximum number of trees and threshold 5 on subset split\item Decision Tree classifier has threshold 2 for instances required at a leaf node\end{enumerate*}. We also evaluate our framework with the probabilistic latent semantic approach but it fails to provide promising results as is the case of Nave Bayes approach as described in Table~\ref{tab:mPreRecalArun}.

\section{Proposed Model}
\label{sec:prop_model}
\begin{figure}[ht!]
\includegraphics[scale=0.6]{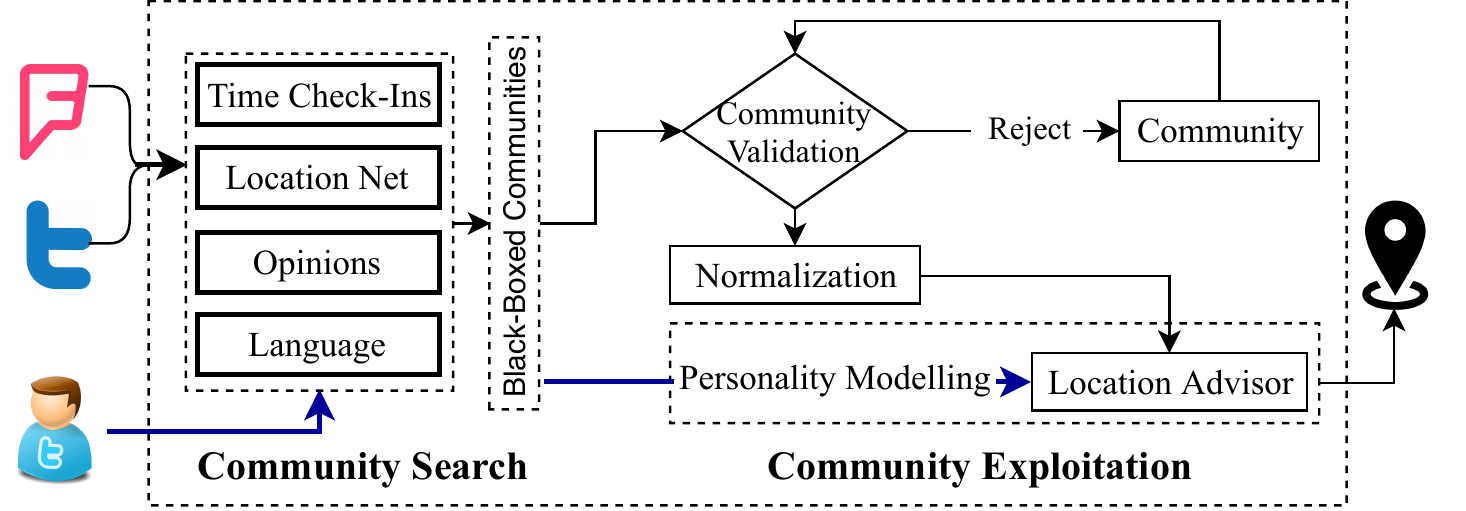}
\caption{Architecture and Methodology of Forecasting Spot.}
\label{fig:Meth_Arch}
\end{figure}

With the advancement of real-time stream processing framework, microblog streams is flooded with random thoughts and opinions of the social user. Considering above statement as fact, we combine the discussed community search and exploitation methods to create a fully-automated system, named as Forecasting Spot and outlined in architecture diagram~\ref{fig:Meth_Arch}. Forecasting Spot takes a Twitter identity as input and in parallel run the \textit{self-timing}, \textit{self-reporting}, \textit{opinions} and \textit{language} based dynamic community search methods. The individual associations comprising of representative clusters returned by each method and collected to black-boxed communities. If there exists a community returned by more than one search method or if a cluster inside community comprising of malformed tuples, the community is returned as a blocked association of tuples and will remain unused for the entire pipeline sessions. The reason for such a decision is that if \begin{enumerate*} \item there is absence of self-reporting information, \item there is absence of self-identification link to location-based social network attributes, \item cluster diversity is less than the threshold (here, 100) and \item absence of having per-day tuples threshold (here, 5) in the cluster \end{enumerate*}. Furthermore, if communities are unable to fulfill specified validations not even for single aspect, thereby strengthening the fact that there is a deficiency of self-reporting positions interlinked with the location-based social network. Hence location will not be advised.
In all other cases, we shortlist the network of like-minded users and clusters inside the network are finely-grained with the presence of user's meta information. Tuples of the finely grained cluster are automatically annotated using Foursquare API which is more reliable than Arun et al. method~\cite{41} and is applied to self-position reporting tuples of the user as well (if available). The finely-grained tuples and most recent tuples of user crawled from their timeline are transformed into vectors using - \textit{personality modeling}. The formation of these vectors is essential as it correlates with self-reporting positions and user personality. Finally, the highly correlated place is predicted using - \textit{Location Advisor} while exploiting contemporary vectors, is then returned.

\section{Experimental Evaluation}
\label{sec:exp_eval}
\subsection{Streams and Metrics Description}
\label{sec:corpus_overview}
\begin{figure}[ht!]
\centering
\includegraphics[scale=0.25]{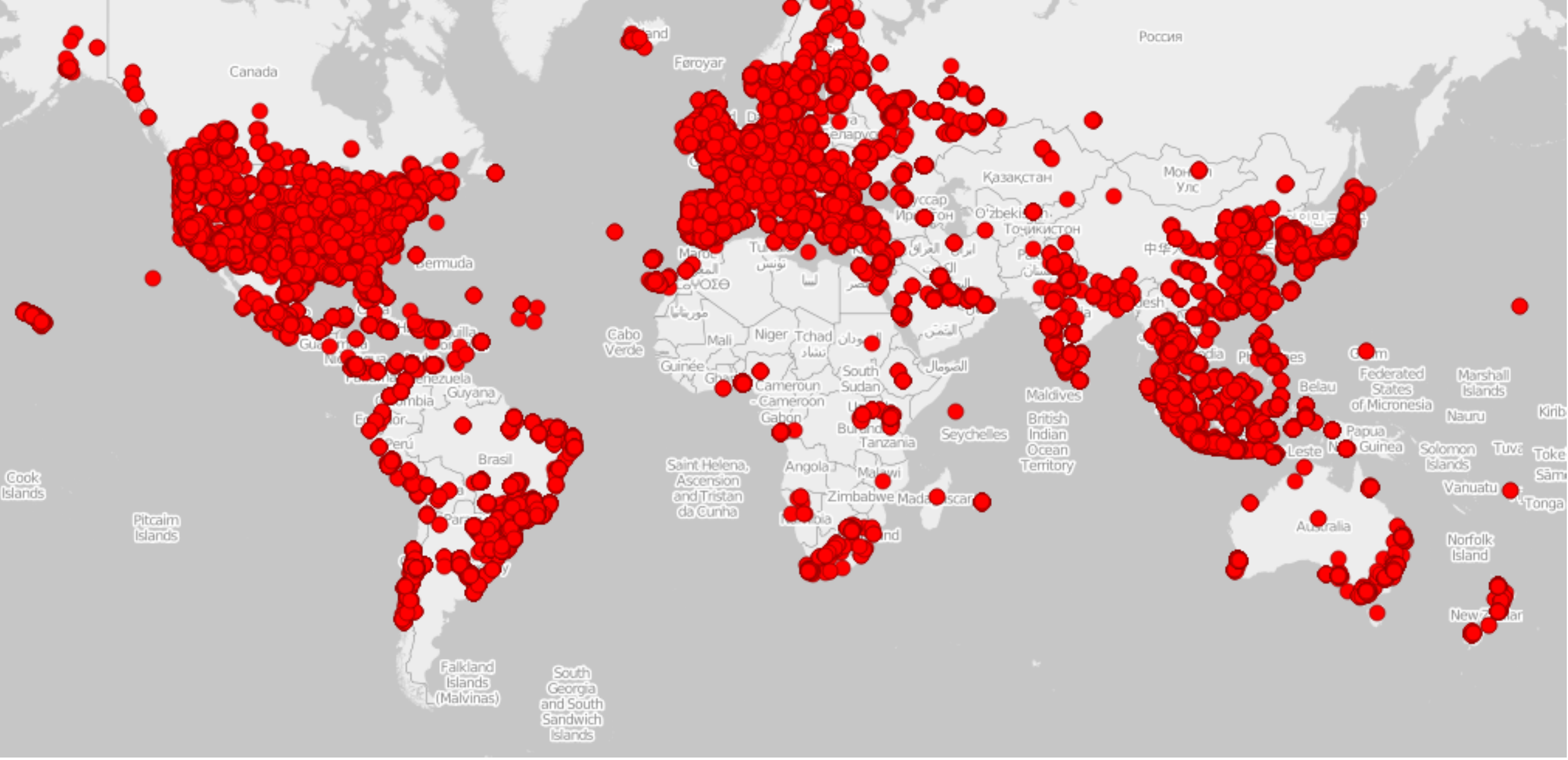}
\caption{Contribution of Self-position reporting in tuples}
\label{fig:corpus}
\end{figure}

We focus on real-time self-position reporting microblog streams subscribed using Twitter streaming API comprising of self-identification link to the location-based social network, \textit{Foursquare}. Towards an extension to the work of authors~\cite{41}, we filter out stream using regular expressions for finding out the mention - \textit{``I'm at''} and \textit{``4sq.com''}. The presence of self-position in tuples is described with geo map as shown in Figure~\ref{fig:corpus} for 1945~(n) users.
\begin{table*}
\centering
\caption{Mean Precision and Recall of Proposed Methods on different algorithms comparing with Ground Truth Observations}
\label{tab:mPreRecalArun}
\begin{tabular}{lcccccccccc}
\hline
\multicolumn{1}{c}{\multirow{2}{*}{Algorithms}} & \multicolumn{2}{c}{Arun et al.\cite{41}} & \multicolumn{2}{c}{Proposed@1}  & \multicolumn{2}{c}{Proposed@2} & \multicolumn{2}{c}{Proposed@3} & \multicolumn{2}{c}{Proposed@4} \\ \cline{2-10} 
\multicolumn{1}{c}{}                            & Precision   & Recall   & Precision      & Recall         & Precision       & Recall       & Precision       & Recall       & Precision       & Recall       \\ \hline
Tree                                            & 0.342       & 0.343    & 0.263          & 0.265          & 0.235           & 0.23         & 0.176           & 0.171        & 0.083           & 0.082        \\
SVM                                             & 0.093       & 0.061    & 0.243          & 0.11           & 0.091           & 0.061        & 0.213           & 0.076        & 0.078           & 0.053        \\
Random Forest                                   & 0.374       & 0.388    & 0.385          & 0.369          & 0.206           & 0.239        & 0.301           & 0.31         & 0.083           & 0.138        \\
\textbf{Neural Network}                         & 0.334       & 0.344    & \textbf{0.621}          & \textbf{0.593}          & 0.144           & 0.153        & 0.481           & 0.483        & 0.085           & 0.119        \\
Naive Bayes                                     & 0.126       & 0.207    & 0.001          & 0.001          & 0.000               & 0.000            & 0.000               & 0.001        & 0.000               & 0.000            \\
Logistic Regression                             & 0.057       & 0.211    & 0.418          & 0.377          & 0.101           & 0.211        & 0.413           & 0.368        & 0.057           & 0.212        \\
\textbf{Overall}                                & 0.221       & 0.259    & \textbf{0.320} & \textbf{0.301} & 0.129           & 0.149        & 0.264           & 0.234        & 0.064           & 0.100        \\ \hline
\end{tabular}
\end{table*}

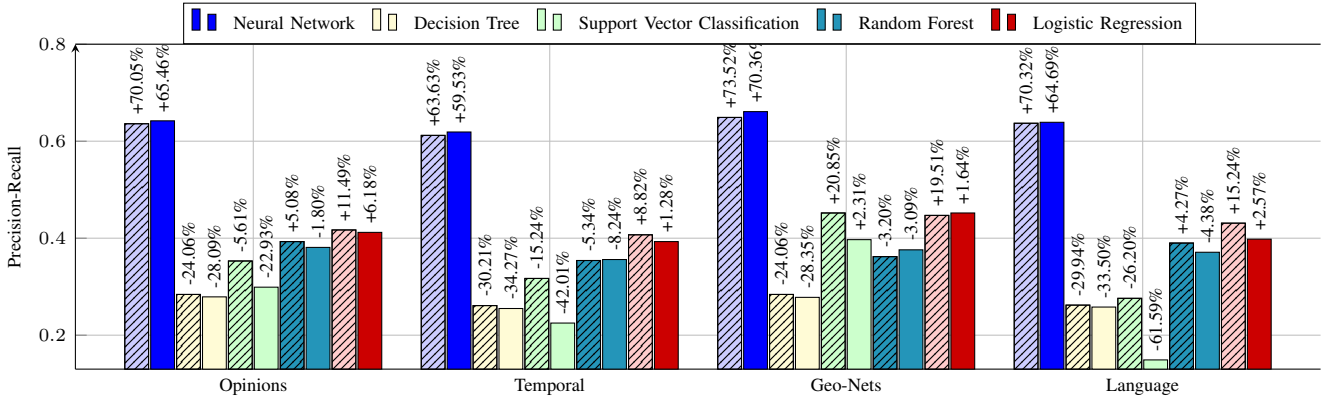
\begin{figure*}
\centering
\begin{tikzpicture}[font=\scriptsize]
    \begin{axis}[
        width  = \textwidth,
		height = 0.25\textheight,
        major x tick style = transparent,
        ybar=2*\pgflinewidth,
        bar width=9pt,
        grid,
		axis y line=left,
        ylabel = {Precision-Recall},
        symbolic x coords={Opinions,Temporal,Geo-Nets,Language},
        xtick = data,
        scaled y ticks = false,
        enlarge x limits=0.20,
        ymin=0.13,
        ymax=0.8,
        tick align=inside,
        xtick=data,
        nodes near coords = \rotatebox{90}{\pgfplotspointmeta},
        nodes near coords align={vertical},
		legend columns=5,
        legend cell align=left,
        legend style={
                at={(0.90,1.00)},
                anchor=south east,
                column sep=1ex,
	            /tikz/every odd column/.append
        },
        point meta=explicit symbolic,
    ]
        \addplot[style={fill=blue!20,mark=none},postaction={
        pattern=north east lines
    }]
            coordinates {
            (Opinions, 0.636)[\scriptsize{+70.05\%}] (Temporal, 0.612)[\scriptsize{+63.63\%}] (Geo-Nets, 0.649)[\scriptsize{+73.52\%}] (Language, 0.637)[\scriptsize{+70.32\%}]
            };
        \addplot[fill=blue,style={mark=none}]
            coordinates {
            (Opinions, 0.642)[\scriptsize{+65.46\%}]
            (Temporal, 0.619)[\scriptsize{+59.53\%}]
            (Geo-Nets, 0.661)[\scriptsize{+70.36\%}]
            (Language, 0.639)[\scriptsize{+64.69\%}]
            };
        
        \addplot[style={fill=yellow!20,mark=none},postaction={
        pattern=north east lines
    }]
            coordinates {
            (Opinions, 0.284)[\scriptsize{-24.06\%}]
            (Temporal, 0.261)[\scriptsize{-30.21\%}]
            (Geo-Nets, 0.284)[\scriptsize{-24.06\%}]
            (Language, 0.262)[\scriptsize{-29.94\%}]
            };
        \addplot[style={fill=yellow!20,mark=none}]
            coordinates {
            (Opinions, 0.279)[\scriptsize{-28.09\%}]
            (Temporal, 0.255)[\scriptsize{-34.27\%}]
            (Geo-Nets, 0.278)[\scriptsize{-28.35\%}]
            (Language, 0.258)[\scriptsize{-33.50\%}]
            };
        
        \addplot[style={fill=green!20,mark=none},postaction={
        pattern=north east lines
    }]
            coordinates {
            (Opinions, 0.353)[\scriptsize{-5.61\%}]
            (Temporal, 0.317)[\scriptsize{-15.24\%}]
            (Geo-Nets, 0.452)[\scriptsize{+20.85\%}]
            (Language, 0.276)[\scriptsize{-26.20\%}]
            };
        \addplot[style={fill=green!20,mark=none}]
            coordinates {
            (Opinions, 0.299)[\scriptsize{-22.93\%}]
            (Temporal, 0.225)[\scriptsize{-42.01\%}]
            (Geo-Nets, 0.397)[\scriptsize{+2.31\%}]
            (Language, 0.149)[\scriptsize{-61.59\%}]
            };
        
        \addplot[style={fill=white!10!cyan!70!black,mark=none},postaction={
        pattern=north east lines
    }]
            coordinates {
            (Opinions, 0.393)[\scriptsize{+5.08\%}]
            (Temporal, 0.354)[\scriptsize{-5.34\%}]
            (Geo-Nets, 0.362)[\scriptsize{-3.20\%}]
            (Language, 0.390)[\scriptsize{+4.27\%}]
            };
        \addplot[style={fill=white!10!cyan!70!black,mark=none}]
            coordinates {
            (Opinions, 0.381)[\scriptsize{-1.80\%}]
            (Temporal, 0.356)[\scriptsize{-8.24\%}]
            (Geo-Nets, 0.376)[\scriptsize{-3.09\%}]
            (Language, 0.371)[\scriptsize{-4.38\%}]
            };
        
        \addplot[style={fill=red!20,mark=none},postaction={
        pattern=north east lines
    }]
            coordinates {
            (Opinions, 0.417)[\scriptsize{+11.49\%}]
            (Temporal, 0.407)[\scriptsize{+8.82\%}]
            (Geo-Nets, 0.447)[\scriptsize{+19.51\%}]
            (Language, 0.431)[\scriptsize{+15.24\%}]
            };
        \addplot[style={fill=red!80!black,mark=none}]
            coordinates {
            (Opinions, 0.412)[\scriptsize{+6.18\%}]
            (Temporal, 0.393)[\scriptsize{+1.28\%}]
            (Geo-Nets, 0.452)[\scriptsize{+1.64\%}]
            (Language, 0.398)[\scriptsize{+2.57\%}]
            };
        \legend{,Neural Network, ,Decision Tree, ,Support Vector Classification, ,Random Forest, ,Logistic Regression}
    \end{axis}
\end{tikzpicture}
\caption{Precision (shaded region) and Recall (dark region) of communities}
\label{fig:neuralCommML}
\end{figure*}
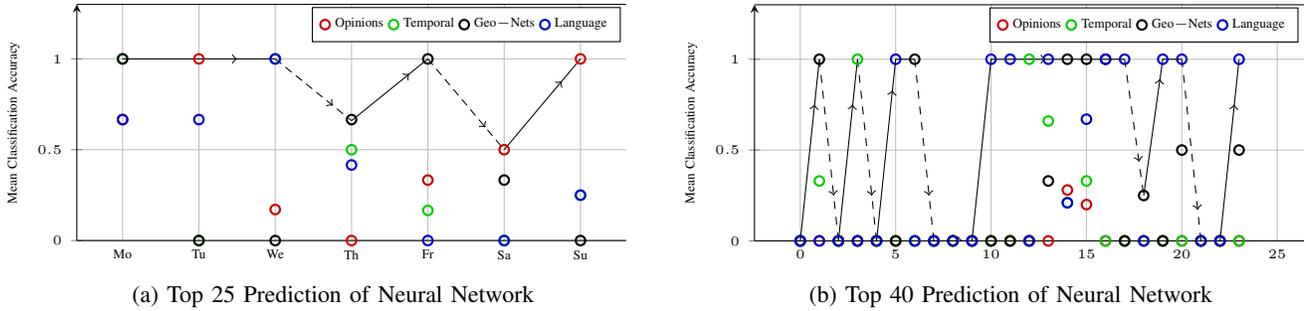
\begin{figure*}
    \begin{subfigure}{0.49\textwidth}
    \begin{tikzpicture}[trim axis right, font=\tiny]
        \begin{axis}[%
            width=\textwidth,
            height = 0.20\textheight,
            symbolic x coords={Mo,Tu,We,Th,Fr,Sa,Su},
            xtick = data,
            scaled y ticks = false,
            enlarge x limits=0.1,
            ymax=1.3,
            major x tick style = transparent,
            axis y line=left,
            ylabel = Mean Classification Accuracy,
            legend columns=5,
            legend cell align=left,
            legend style={
                    at={(0.99,0.84)},
                    anchor=south east,
                    column sep=1ex,
                    /tikz/every odd column/.append
            },
            legend style={column sep=0.00001cm, text depth=0pt},
            grid,
        	scatter/classes={%
            	a={mark=o,draw=black, red!80!black, thick, mark size=1.8,fill=none},
            	b={mark=o,draw=black, green!80!black, thick, mark size=1.8,fill=none},
            	c={mark=o,draw=black, black, thick, fill=none, mark size=1.8},
            	d={mark=o,draw=black, blue!80!black, thick, mark size=1.8,fill=none}
                }]
        \addplot[scatter,only marks,scatter src=explicit symbolic]%
            table[meta=label] {
                x y label
                Mo	0.666	a	
                Tu	1.000	a
                We	0.171	a
                Th	0.000	a
                Fr	0.333	a
                Sa	0.500	a
                Su	1.000	a
                
                Mo	1.000	b
				Tu	0.000	b
				We	1.000	b
				Th	0.500	b
				Fr	0.166	b
                Sa	0.000	b
                Su	0.250	b
                
                Mo	1.000	c
                Tu	0.000	c
                We	0.000	c
                Th	0.666	c
                Fr	1.000	c
                Sa	0.333	c
                Su	0.000	c
                
                Mo	0.666	d
                Tu	0.666	d
                We	1.000	d
                Th	0.416	d
                Fr	0.000	d
                Sa	0.000	d
                Su	0.250	d
            };
        \begin{scope}[very thick,decoration={
          markings,
          mark=at position 0.75 with {\arrow{>}}}
          ] 
          \draw[thin,postaction={decorate}] (axis cs:Mo,1.000) -- node[left]{} (axis cs:Tu,1.00) -- node[left]{} (axis cs:We,1.00);
          \draw[thin,densely dashed,postaction={decorate}] (axis cs:We,1.00) -- node[left]{} (axis cs:Th,0.66);
          \draw[thin,postaction={decorate}] (axis cs:Th,0.66) -- node[left]{} (axis cs:Fr,1.00);
          \draw[thin,densely dashed,postaction={decorate}] (axis cs:Fr,1.00) -- node[left]{} (axis cs:Sa,0.500);
          \draw[thin,postaction={decorate}] (axis cs:Sa,0.500) -- node[left]{} (axis cs:Su,1.00);
        \end{scope}
        \legend{Opinions,Temporal,Geo$-$Nets,Language}
        \end{axis}
    \end{tikzpicture}    
    \caption{Top 25 Prediction of Neural Network}
    \label{fig:SpotW}
    \end{subfigure}
    \begin{subfigure}{0.49\textwidth}
    \begin{tikzpicture}[trim axis right, font=\tiny]
        \begin{axis}[%
            width=\textwidth,
            height = 0.20\textheight,
            scaled y ticks = false,
            enlarge x limits=0.1,
            ymax=1.3,
            xmax=24,
            major x tick style = transparent,
            axis y line=left,
            ylabel = Mean Classification Accuracy,
            legend columns=5,
            legend cell align=left,
            legend style={
                    at={(0.99,0.84)},
                    anchor=south east,
                    column sep=1ex,
                    /tikz/every odd column/.append
            },
            legend style={column sep=0.00001cm, text depth=0pt},
            grid,
        	scatter/classes={%
            	a={mark=o,draw=black, red!80!black, thick, mark size=1.8,fill=none},
            	b={mark=o,draw=black, green!80!black, thick, mark size=1.8,fill=none},
            	c={mark=o,draw=black, black, thick, fill=none, mark size=1.8},
            	d={mark=o,draw=black, blue!80!black, thick, mark size=1.8,fill=none}
                }]
        \addplot[scatter,only marks,scatter src=explicit symbolic]%
            table[meta=label] {
                x y label
                0	0.00 a
                1	0.00 a
                2	0.00 a
                3	0.00 a
                4	0.00 a
                5	0.00 a
                6	0.00 a
                7	0.00 a
                8	0.00 a
                9	0.00 a
                10	0.00 a
                11	0.00 a
                12	0.00 a
                13	0.00 a
                14	0.28 a
                15	0.20 a
                16	0.00 a
                17	0.00 a
                18	0.00 a
                19	0.00 a
                20	0.00 a
                21	0.00 a
                22	0.00 a
                23	0.00 a
                
                15	0.33 b
                22	0.00 b
                19	0.00 b
                8	0.00 b
                0	0.00 b
                17	0.00 b
                23	0.00 b
                6	0.00 b
                12	1.00 b
                9	0.00 b
                20	0.00 b
                11	0.00 b
                1	0.33 b
                14	0.21 b
                18	0.00 b
                21	0.00 b
                4	0.00 b
                7	0.00 b
                2	0.00 b
                10	0.00 b
                3	1.00 b
                16	0.00 b
                5	0.00 b
                13	0.66 b
                
                15	1.00 c
                22	0.00 c
                19	0.00 c
                8	0.00 c
                0	0.00 c
                17	0.00 c
                23	0.50 c
                6	1.00 c
                12	0.00 c
                9	0.00 c
                20	0.50 c
                11	0.00 c
                1	1.00 c
                14	1.00 c
                18	0.25 c
                21	0.00 c
                4	0.00 c
                7	0.00 c
                2	0.00 c
                10	0.00 c
                3	0.00 c
                16	1.00 c
                5	0.00 c
                13	0.33 c
                
                15	0.67 d
                22	0.00 d
                19	1.00 d
                8	0.00 d
                0	0.00 d
                17	1.00 d
                23	1.00 d
                6	0.00 d
                12	0.00 d
                9	0.00 d
                20	1.00 d
                11	1.00 d
                1	0.00 d
                14	0.21 d
                18	0.00 d
                21	0.00 d
                4	0.00 d
                7	0.00 d
                2	0.00 d
                10	1.00 d
                3	0.00 d
                16	1.00 d
                5	1.00 d
                13	1.00 d

            };
        \begin{scope}[very thick,decoration={
          markings,
          mark=at position 0.75 with {\arrow{>}}}
          ]           
          \draw[thin,postaction={decorate}] (axis cs:0,0.000) -- node[left]{} (axis cs:1,1.00);
          \draw[thin,postaction={decorate}, dashed] (axis cs:1,1.00) -- node[left]{} (axis cs:2,0.00);
          \draw[thin,postaction={decorate}] (axis cs:2,0.000) -- node[left]{} (axis cs:3,1.00);
          \draw[thin,postaction={decorate}, dashed] (axis cs:3,1.00)-- node[left]{} (axis cs:4,0.00);
          \draw[thin,postaction={decorate}] (axis cs:4,0.000) -- node[left]{} (axis cs:5,1.00) -- node[left]{} (axis cs:6,1.00);
          \draw[thin,postaction={decorate}, dashed] (axis cs:6,1.00) -- node[left]{} (axis cs:7,0.00);
          \draw[thin,postaction={decorate}] (axis cs:7,0.000) -- node[left]{} (axis cs:8,0.00) -- node[left]{} (axis cs:9,0.00);
          \draw[thin,postaction={decorate}] (axis cs:9,0.00) -- node[left]{} (axis cs:10,1.00) -- node[left]{} (axis cs:11,1.00) -- node[left]{} (axis cs:12,1.00) -- node[left]{} (axis cs:13,1.00) -- node[left]{} (axis cs:14,1.00) -- node[left]{} (axis cs:15,1.00) -- node[left]{} (axis cs:16,1.00) -- node[left]{} (axis cs:17,1.00);
          \draw[thin,postaction={decorate}, dashed] (axis cs:17,1.00) -- node[left]{} (axis cs:18,0.25);
          \draw[thin,postaction={decorate}] (axis cs:18,0.25) -- node[left]{} (axis cs:19,1.00) -- node[left]{} (axis cs:20,1.00);
          \draw[thin,postaction={decorate}, dashed] (axis cs:20,1.00) -- node[left]{} (axis cs:21,0.00) -- node[left]{} (axis cs:22,0.00);
          \draw[thin,postaction={decorate}] (axis cs:22,0.00) -- node[left]{} (axis cs:23,1.00);

        \end{scope}
        \legend{Opinions,Temporal,Geo$-$Nets,Language}
        \end{axis}
    \end{tikzpicture}    
    \caption{Top 40 Prediction of Neural Network}
    \label{fig:spotM}
    \end{subfigure}
    \caption{Forecasting Spot - Weekly and Hourly Prediction Performance}
    \label{fig:SpotWeeklyandHourly}
\end{figure*}
The performance of proposed framework is delineated with three evaluation metrics running on top of keras 2.2.0 with tensorflow back-end on a colaboratory instance having Nvidia Tesla K80 acceleration. These metrics are indexed as:
\paragraph{\textbf{Mean Classification Accuracy:}} The accuracy of a multi-class recommendation framework for $2^{n+1}$ communities having $2^{n_{i}+1}$ clusters where $n_{i}$ users available in community ($M_{\alpha}^{A}$) is defined as ratio of identities for whom true positive spot is recommended ($T_{\alpha}^{k})$) and users ($n_{i}^{j}$) for whom location is fore-casted belongs to $C_{i}$ cluster. It measures the effectiveness of framework in recommending highly correlated spot corresponds to user identity. Higher the accuracy, better is the framework. Formally, mean classification accuracy is given by:
\begin{equation}
\resizebox{.8\hsize}{!}{$f(\alpha_1,\alpha_2,\ldots,\alpha_{n})=\sum_{i=\tiny{M_{\alpha}^{A}}}^{2^{n+1}} \cdot \sum_{j=C_{i}}^{2^{n_{i}+1}} \left\lbrace \frac{1}{\sum_{k=1}^{n_{i}^{j}}(\alpha_{k})} \cdot \left( \sum_{k=1}^{n_{i}^{j}} (T_{\alpha}^{k}) \right) \right\rbrace $}
\label{eq:mca}
\end{equation}

\paragraph{\textbf{Mean Average Precision-Recall:}} The precision of a multi-class recommendation framework for $2^{n+1}$ communities having $2^{n_{i}+1}$ clusters where $n_{i}$ users available in community ($M_{\alpha}^{A}$) is defined as:
\begin{equation}
\resizebox{.85\hsize}{!}{$ f(\alpha_1,\alpha_2,\ldots,\alpha_{n}) = \sum_{i=\tiny{M_{\alpha}^{A}}}^{2^{n+1}} \cdot \sum_{j=C_{i}}^{2^{n_{i}+1}} \left\lbrace \frac{1}{\sum_{k=1}^{n_{i}^{j}}(\alpha_{k})} \cdot \left( \sum_{k=1}^{n_{i}^{j}} PR(\alpha_{k})\ \times\ Rel(\alpha_{k}) \right) \right\rbrace $}
\label{eq:mapr}
\end{equation}
while PR resembles the precision of user ($\alpha_{k}$) for whom spot is recommended, boolean \textit{Rel} denotes the relativeness of user ($\alpha_{k}$) to resembles spot is correct and $n_{k}$ depicts the subset of users, belongs to $C_{i}$ cluster for whom spot is forecasted. It measures the efficacy of framework in recommending highly correlated spot corresponds to user identity. Higher the precision and recall, better is the rank at which correct spot is returned from communities.

\subsection{Results and Inferences}
We measured the efficacy of \textit{Forecasting Spot} by performing training on randomly selected 80\% of users and evaluation on the remaining while ensuring the presence of at least 30 days check-in history for 20\% of users. We have used n fold cross-validation (here, 10) method for initial ranking of proposed methods at training phase and later cross-verified at testing phase. We observed that \textit{Neural Network} emerged as a de-facto multi-label recommendation framework while comparing with state-of-the-art observations as shown in Table~\ref{tab:mPreRecalArun} relatively having high precision with high recall. Table~\ref{tab:mPreRecalArun} list the performance of vector space exploitation and non-vector space feature exploitation to state-of-the-art contributions. However, we analyzed that distributed representation of tuples outperforms the vector space of word embeddings resulting in the overall precision of 64.2\% for the framework. We further evaluate our framework while discarding the personality features to strengthen the fact that they have a high correlation with self-reporting locations along with a distributed representation of tuples beside promising improvement of 33.47\% and 33.54\% in precision and recall respectively.
We describe the performance of individual communities discussed in Section~\ref{sec:comm_search_method} over widely used machine learning algorithms as shown in Figure~\ref{fig:neuralCommML}. We observed that \textit{Geo-Nets} community have the maximum contribution with a performance gain of 73.52\% in precision and 70.36\% in recall respectively while comparing with traditional methods used in literature. Simultaneously, the \textit{Temporal} community has the least contribution with a performance gain of 63.63\% in precision and 59.53\% in recall respectively. The traditional methods have better results with logistic regression and naive bayes algorithms. Due to their complexities~\cite{32}, these algorithms perform worst towards large data streams as shown in Table~\ref{tab:mPreRecalArun} with the contribution of 0.057\% and 0.126\% precision respectively. However, towards an initial baseline improvement, the neural network evolve as a fruitful asset with an increment in precision (4.85\% and 3.64\% respectively) on large data streams. Therefore, with an evaluation on 20\% of users, the neural network provides promising results due to the presence of only likely minded users as an input features as shown in Figure~\ref{fig:neuralCommML}. For example, users with same native language always tend to visit similar places. Therefore, the \textit{Location Advisor} methods exploit feed-forward neural network model in forecasting locations. 

We described the training accuracies of forecasting spot communities in Table~\ref{tab:trainPrRe}. We observed that ``cluster 1'' of community \textit{Geo-Nets} comprises 6.25\% users but still, it retains maximum performance with 0.687 F1-Score. Although ``South America'' and ``Australia'' clusters comprise a similar set of users (2.56\%) still it scores differently with 0.642 and 0.643 precisions respectively for \textit{Language} community. Although, ``Antarctica'' cluster achieves maximum precision (0.644) comprising 69.79\% users however it's precision almost equivalent to ``South America''. The ``Night'' cluster sustain maximum F1-Score (0.648) in \textit{Temporal} community with 19.25\% of users in total. we achieve maximum F1-Score of 0.641 in recommending a spot to 75.83\% users having ``Neutral'' opinions towards their self-reporting places.
We depict the weekly and hourly recommendation performance of Forecasting Spot framework in Figure~\ref{fig:SpotWeeklyandHourly}. The weekly forecast of a user clustered with simultaneous 25 predictions is delineated in Figure~\ref{fig:SpotW} while ensuring negligence of the user in training corpus. We observed that ``Geo-Nets'' and ``Opinions'' communities are actively contributing towards the strengthening of overall performance of the framework. The mean contribution of remaining communities is significantly degrading over the weekdays. However, the ``Geo-Nets'' still maintains 66\% of mean classification accuracy for ``Thursday'' cluster and 50\% of mean classification accuracy is attained by ``Opinions'' community on ``Saturday'' cluster. Similarly, the long-range hourly recommendation of the user spot is delineated in Figure~\ref{fig:spotM} for simultaneous 50 predictions while ensuring negligence of the user in training corpus. We observed that ``Geo-Nets'', ``Language'' and ``Temporal'' communities are actively contributing in the strengthening of the overall performance of the framework. Although the mean contribution of remaining community is degrading. However, the ``opinion'' community still retain 28\% and 20\% of mean classification accuracy on 1400 hrs and 1500hrs respectively. We observe that there is steep increment or decrement in performance of communities with a change in clustering scenario. For example, on weekends user generally prefer hangout to normal days. However, our framework automatically adapts to the change in environment to recommend the most reliable spot.

\section{Conclusions and Future Work}
\label{sec:conclusion}
\begin{table}
\centering
\caption{Mean and Standard Deviation over Communities}
\label{tab:trainPrRe}
\resizebox{0.8\hsize}{!}{
\begin{tabular}{llccc}
\hline
\multicolumn{1}{c}{Community}                   & \multicolumn{1}{c}{Label} & F1-Score               & Precision              & Recall                 \\ \hline \hline
\multicolumn{1}{l|}{\multirow{10}{*}{Geo-Nets}} & Cluster 1                 & \textbf{0.687}                  & 0.684                  & 0.692                  \\
\multicolumn{1}{l|}{}                           & Cluster 2                 & 0.628                  & 0.627                  & 0.630                   \\
\multicolumn{1}{l|}{}                           & Cluster 3                 & 0.639                  & 0.634                  & 0.645                  \\
\multicolumn{1}{l|}{}                           & Cluster 4                 & 0.658                  & 0.648                  & 0.669                  \\
\multicolumn{1}{l|}{}                           & Cluster 5                 & 0.674                  & 0.669                  & 0.681                  \\
\multicolumn{1}{l|}{}                           & Cluster 6                 & 0.598                  & 0.606                  & 0.591                  \\
\multicolumn{1}{l|}{}                           & Cluster 7                 & 0.681                  & 0.672                  & 0.691                  \\
\multicolumn{1}{l|}{}                           & Cluster 8                 & 0.667                  & 0.644                  & 0.692                  \\
\multicolumn{1}{l|}{}                           & Cluster 9                 & 0.671                  & 0.665                  & 0.679                  \\
\multicolumn{1}{l|}{}                           & Cluster 10                & 0.644                  & 0.642                  & 0.647                  \\ \hline
\multicolumn{1}{l|}{\multirow{7}{*}{Language}}  & Africa                    & 0.636                  & 0.635                  & 0.638                  \\
\multicolumn{1}{l|}{}                           & Antarctica                & \textbf{0.645}                  & 0.644                  & 0.647                  \\
\multicolumn{1}{l|}{}                           & Asia                      & 0.630                  & 0.629                  & 0.632                  \\
\multicolumn{1}{l|}{}                           & Europe                    & 0.627                  & 0.627                  & 0.629                  \\
\multicolumn{1}{l|}{}                           & North America             & 0.641                  & 0.641                  & 0.643                  \\
\multicolumn{1}{l|}{}                           & South America             & 0.643                  & 0.642                  & 0.645                  \\
\multicolumn{1}{l|}{}                           & Australia                 & 0.643                  & 0.643                  & 0.645                  \\ \hline
\multicolumn{1}{l|}{\multirow{4}{*}{Temporal}}  & Morning                   & 0.618                  & 0.616                  & 0.621                  \\
\multicolumn{1}{l|}{}                           & Afternoon                 & 0.635                  & 0.633                  & 0.639                  \\
\multicolumn{1}{l|}{}                           & Evening                   & 0.606                  & 0.602                  & 0.612                  \\
\multicolumn{1}{l|}{}                           & Night                     & \textbf{0.648}                  & 0.698                  & 0.605                  \\ \hline
\multicolumn{1}{l|}{\multirow{3}{*}{Opinions}}  & Positive                  & 0.640                  & 0.637                  & 0.644                  \\
\multicolumn{1}{l|}{}                           & Negative                  & 0.635                  & 0.631                  & 0.640                   \\
\multicolumn{1}{l|}{}                           & Neutral                   & \textbf{0.641}                  & 0.640                   & 0.644                  \\ \hline
\multicolumn{1}{l|}{\textbf{Overall}}           & \textbf{}                 & \textbf{0.643 (0.021)} & \textbf{0.642 (0.022)} & \textbf{0.645 (0.026)} \\ \hline
\end{tabular}
}
\end{table}
In this paper, we propose a methodology to address the problem recommending a plausible place to the visit through microblog streams. We exploit traditional approach towards wide applicability to the distinct social network. In such scenarios, fruitful information is submerged with various unknown locations. However, we identified self-location mentioned in the stream tuples helps in automatically correlating with the location-based social network and providing aid towards annotation with location category. We provide a formal definition, algorithmic methods, and dynamic validations for the formation of communities using social network streams. We reveal that combination of various community search methods - Dynamic Opinion Search, Dynamic Language Search, Dynamic Geo-Nets Search and Dynamic Temporal Search helps in exploiting fluctuating point-of-interest of contributing users towards recommending the correct place having in-depth granularity.  Hence, our fully automated framework ``Forecasting Spot'' integrate these tuples into the multilayer feed-forward neural network using vector space embedding of active community resembles user with a high confidence score. The framework achieves 62.10 \% mean average precision on 1,20,000 embeddings of unlabeled users and surprisingly, 85.92 \% increment to state-of-the-art approach. We plan to extend our contributions to include various methods utilizing static contents e.g gender-based communities which may further contribute to enhancing the precision, recall, and accuracy of the framework. Hence, \textit{Forecasting Spot} will provide a similar promising result with streams of another network as well.



\bibliographystyle{IEEEtran}
\bibliography{cikm_ref}

\end{document}